\documentclass[11pt,a4paper]{article}
\usepackage[top=2.3cm,right=2.3cm,left=2.3cm,bottom=2.3cm]{geometry}
\usepackage{bbold}
\usepackage{cite}
\usepackage[pdfstartview=FitH,colorlinks=true,linkcolor=black,anchorcolor=black,citecolor=black,urlcolor=black]{hyperref}
\usepackage[english]{babel}
\usepackage{amsmath,amssymb,titling,authblk}
\usepackage{amsmath,slashed}
\usepackage{cite}
\usepackage{xcolor}

\definecolor{verde}{cmyk}{.83,.21,1,.08}
\definecolor{darkorchid}{rgb}{0.6, 0.2, 0.8}

\newcommand{\ii}{\mathrm{i}}
\newcommand{\bM}{\partial\mathcal{M}}
\newcommand{\dd}{\mathrm{d}}

\newcommand{\be}{\begin{equation}}
\newcommand{\ee}{\end{equation}}
\newcommand{\bea}{\begin{eqnarray}}
\newcommand{\eea}{\end{eqnarray}}

\newcommand{\la}{\label}

\begin{document}
\setlength{\droptitle}{-6pc}
\pretitle{\begin{flushright}\small
\end{flushright}\vspace*{2pc}%
\begin{center}\LARGE}
\posttitle{\par\end{center}}

\title{Remark on the  synergy between \\[3pt] the heat kernel techniques\\[3pt] and the parity anomaly.
\vspace{5pt}
}

\renewcommand\Affilfont{\itshape}
\setlength{\affilsep}{1.5em}
\renewcommand\Authands{ and }

\author[1,2]{Maxim Kurkov\thanks{max.kurkov@gmail.com}}
\author[3]{Lorenzo Leone\thanks{lorenzoleone95@gmail.com}}
\affil[1]{Dipartimento di Fisica ``Ettore Pancini'', Universit\`{a} di Napoli {\sl Federico II}, \newline  Complesso Universitario di Monte S. Angelo, Via Cintia, I-80126 Napoli, Italy \vspace{5pt}}
\affil[2]{INFN, Sezione di Napoli, Via Cintia, 80126 Napoli, Italy\vspace{5pt}}
\affil[3]{Department of Physics, University of Massachusetts Boston, Massachusetts 02125, USA}

\date{}

\maketitle 

\begin{abstract}\noindent
In this paper, we demonstrate that not only the heat kernel techniques are useful for computation of the parity anomaly, but also the parity anomaly turns out to be a powerful mean in studying the heat kernel.  We show that the gravitational parity anomaly on 4D manifolds with boundaries can be calculated using the general structure of the heat kernel coefficient $a_5$ for mixed boundary conditions, keeping all the weights of various geometric invariants as unknown numbers. The  symmetry properties of the $\eta$-invariant allow to fix all the relevant unknowns. As a byproduct of this calculation we get an efficient and independent crosscheck (and confirmation) of the correction of the general structure of $a_5$ for mixed boundary conditions, previously suggested in Ref.~\cite{Moss:2012dp}.
\end{abstract}

\newpage
\section{Introduction.}
The heat kernel techniques exhibit a broad variety of applications to Quantum Field Theory (QFT) as far as the one-loop effective action is concerned: ultraviolet divergences, chiral and conformal anomalies as well as various asymptotics of the effective action can be studied using the heat-kernel methods \cite{manual,Fursaev:2011zz}. These techniques play a crucial role in the spectral approach to noncommutative 
geometry~\cite{Connes1997,WalterBook,AllThat}, allowing to extract a physically relevant information from the 
 bosonic spectral action and, in particular, to calculate the mass of the Higgs boson in this formalism~\cite{HiggsMass}.

A crucial role is played by the Seeley-De Witt (or simply heat kernel) coefficients, associated to a  Laplace-type operator, which acts on smooth sections of a given bundle over a given manifold.
 Each heat kernel coefficient appears to be a 
linear combination of  integrals of
  local bulk and boundary geometric invariants\footnote{
Upon the ``bulk or boundary geometric invariants" we mean the objects which
 are independent on the particular local frame for the bundle 
 and local coordinate systems on 
the manifold and and its boundary respectively.}, which have correct canonical mass dimensions, see  Sec.~2 for details.
An 
important property of the heat kernel coefficients is their \emph{universality} \cite{manual}: the weights of various geometric invariants, which enter in the heat kernel coefficients, 
do not depend on a particular choice of the Riemannian manifold, 
 the  Laplace-type operator  
and  the quantities 
which define the boundary conditions\footnote{Throughout this article we are talking about the mixed boundary conditions only, which are defined by two complementary projectors $\Pi_{\pm}$ and one endomorphism $S$, see the details in Sec.~2.}. Moreover, these weights depend on a dimension $n$ of the manifold  in a very simple way: via a common factor of $1/(4\pi)^{n/2}$ for the bulk invariants and via a common factor of $1/(4\pi)^{(n-1)/2}$
for the boundary ones. 
This universality explains a peculiar role played by particular calculations e.g. on a ball~\cite{ball1,ball2} for a simple Laplacian with convenient boundary conditions: constrains on the wights, which come out from particular examples are valid always. In order to find a general
formula for the coefficient $a_k$ the following strategy is usually used~\cite{Branson:1990xp,Branson:1995cm,Kirsten:1997qd,a5}.
\begin{itemize}
\item{First, one has to write down all possible independent invariants, allowed by the dimensional analysis, with unknown weights.} 
\item{After that, generating constrains from particular calculations on a ball and/or using other tools, such as a method of conformal variations  
one has to determine all the unknowns.}
\end{itemize}
Unfortunately the number of independent geometric invariants grows up very fast with the grows of the number $k$ of the heat kernel coefficient.  
Already at $k=5$
this number exceeds 150, see~\cite{a5}. Therefore a generation of a sufficient number of constraints on the undetermined weights becomes a complicated task.  In this article we demonstrate that the \emph{parity anomaly},  a purely QFT object, which, however, exhibits nontrivial connections with the heat kernel techniques, can be useful at this point.

Dynamical breakings of symmetries have been considered in various physical contexts using various mechanisms~\cite{Kurkov:2016zpd,Jones:2017ejm,Ferreira:2018itt,Ferreira:2018qss,Liang:2019fkj,Ferreira:2019ywk}.
 The parity anomaly, see~\cite{Niemi:1983rq,Redlich:1983dv},  was introduced as a dynamical breaking of parity due to quantum corrections in theories of three-dimensional fermions, which interact with external gauge fields and which are parity-invariant at the classical level. It was demonstrated that in the gauge-invariant setup the one-loop effective action necessarily contains a Chern-Simons term, which violates parity.
In~\cite{AG} the notion of the parity anomaly was generalised to odd-dimensional fermions interacting with gravitational backgrounds, and
the parity anomaly was  related to the $\eta$-invariant, which describes the spectral asymmetry of the fermionic Dirac operator. 
The gravitational contribution to the parity anomaly in three dimensions was computed in~\cite{Goni:1986cw,Vuorio:1986ju,vanderBij:1986vn}, and it is given by the gravitational Chern-Simons term\footnote{This object,  originally introduced in~\cite{Deser1,Deser2},
exhibits interesting applications and generalisations in the higher-spin physics~\cite{Grigoriev:2019xmp}.}. 
Manifolds with boundaries\footnote{Theories with boundaries naturally arise in the context of BCFT (for the recent progress see e.g.~\cite{Chu:2018ntx,Miao:2018qkc,Miao:2018dvm,Chu:2018fpx,Chu:2019rod,Zheng:2019xeu} and refs therein), and in the noncommutative geometry \cite{Lizzi:2006bu}.} were considered in the context of the parity anomaly (and the $\eta$-invariant) in~\cite{Witten:2015aba} upon the APS boundary conditions and in \cite{Parity,Gparity,ChS,Witten:2019bou} upon the local bag boundary conditions.  Nowadays these parity-odd effects receive significant attention in  condensed matter physics~\cite{Golan:2018tdy,Golan:2019svj,Wang:2017xhg,Fialkovsky:2018fpo,Vassilevich:2019mhl} as well as in various areas of QFT and mathematical physics~\cite{Vassilevich:2018aqu,Lapa:2019fiv,Alonso-Izquierdo:2019tms,MateosGuilarte:2019eem,Palumbo:2019pvp}.

 In \cite{Parity,Gparity} the heat kernel techniques were used in order to calculate gauge and gravitational contributions to the parity anomaly on four-dimensional manifolds with boundaries, relating it
to the third and the fifth heat kernel coefficients.   
In the present paper we  obtained a very interesting result which reflects the synergy between the heat kernel and the parity anomaly. What has emerged is that not only the heat kernel techniques are useful for computation of the parity anomaly, but also the parity anomaly turns out to be a powerful mean in studying the heat kernel.

We demonstrate that the gravitational contribution to the parity anomaly can be calculated using the general structure of the fifth heat kernel coefficient, keeping 
all the weights of various geometric invariants as unknown numbers. The symmetry properties of the $\eta$-invariant allow to fix 
all the relevant unknowns.

This article is organised as follows. In Sec.~2 we briefly describe the relevant aspects of the heat kernel expansion, without going too deep into technical details. In Sec.~3 we review the connection between the parity anomaly and the heat kernel techniques.  Sec.~4 is devoted to the new results, announced above.  
Notations, conventions and various technicalities are collected in the Appendices {\bf A} - {\bf D}.

\section{Heat kernel techniques.}
Let $V$ be a vector bundle over a Riemannian $n$-dimensional manifold $\mathcal{M}$.
A Laplace-type operator $L$ is the second-order differential operator, which acts on smooth sections of $V$, and which has the following form\footnote{Note, that the combination $g^{\mu \nu} \nabla_{\mu}\nabla_{\nu}$ contains the Christoffel connection $\Gamma_{\mu\nu}^\rho = \frac{g^{\rho\lambda}}{2}\left(- \partial_{\lambda}g_{\mu\nu}
 +\partial_{\mu}g_{\nu\lambda} + \partial_{\nu}g_{\lambda\mu} \right)$ as well,  since the covariant derivative $\nabla_{\mu}$ acts on the object which carries one world index $\nu$. }:
\begin{align}
L=-g^{\mu \nu} \nabla_{\mu}\nabla_{\nu} + E, \quad\mbox{with}\quad \nabla_{\nu}\equiv\partial_{\nu} + \omega_{\nu},
\label{25}
\end{align}
where $g_{\mu \nu}$ is the Riemannian metric on $\mathcal{M}$, $\omega$ is a connection on $V$ and the matrix valued function $E$ (in bundle indices) is an endomorphism of $V$. 
 From now on we assume the manifold $\mathcal{M}$ to be compact and with a smooth boundary $\partial\mathcal{M}$. Local coordinates on $\mathcal{M}$ and $\partial\mathcal{M}$  are denoted through $x^{\mu}$, $\mu = 1,...,n$ and $\tilde{x}^j$, $j = 1,..,n-1$ respectively, 
 and in what follows $\tilde{g}_{jk}$ stands for the induced metric on the boundary. A complete summary of the notations and conventions is presented in the Appendix {\bf A}.

By definition the \emph{heat operator} $e^{-t L}$ maps a square integrable section $f$ of $V$ to the solution  $u := e^{-t L} f$ of the
 initial-boundary value problem for the heat equation:
\be
\partial_t u = - L u,\quad u_{t = 0} = f, \quad \mathcal{B}u |_{\partial \mathcal{M}} =0, \la{iBVP}
\ee
where the boundary operator $\mathcal{B}$, defined on $V|_{\partial\mathcal{M}}$, specifies the boundary conditions.
In what follows we restrict ourselves to the mixed boundary conditions:
\be 
\mathcal{B} = \Pi_{-} +(\nabla_n + S)\Pi_{+},  \la{gbc}
\ee
where $\Pi_{+}$ and $\Pi_{-}$ are two complementary projectors defined on $V|_{\partial\mathcal{M}}$, $S$ is a matrix-valued function defined on $\partial\mathcal{M}$ and
$\nabla_n \equiv n^{\mu} \nabla_{\mu}$ is the covariant derivative along the unit inward pointing normal $n$ to the boundary.

For a given matrix valued function 
$Q_0$ the \emph{heat trace} 
is defined as follows:
\be
K(Q_0, L, \mathcal{B},t) = \mathrm{Tr}_{L^2}\left( Q_0 e^{-t L}\right), \la{hc0}
\ee
and at  $t\longrightarrow 0^+$ the  asymptotic heat kernel expansion takes place:
\be
K(Q_0,L,\mathcal{B},t) \simeq \sum_{k=0}^{\infty} t^{\frac{k-n}{2}} a_k(Q_0,L,\mathcal{B}).
\ee
The quantities $a_n$ are called the Seeley-De Witt (or simply heat kernel) coefficients. These coefficients have
the structure:
\be
a_k(Q_0,L,\mathcal{B}) = \int_{\mathcal{M}} \dd x \sqrt{g}\, a^{\mathcal{M}}_k(x,Q_0,L) 
+ \int_{\partial\mathcal{M}} \dd \tilde{x} \sqrt{\tilde{g}}\, a^{\partial\mathcal{M}}_k (\tilde{x},Q_0,L,\mathcal{B}).
\ee

Each bulk density  $a^{\mathcal{M}}_k(x,Q_0,L)$ is given by a linear combination of local bulk geometric invariants of the canonical mass dimension\footnote{The  endomorphism $Q_0$ is assumed to be dimensionless. For other canonical mass dimensions in this context see e.g. Sec. 4.4 of \cite{Fursaev:2011zz}.} $k$, 
constructed from the endomorphisms $E$ and $Q_0$, the metric $g_{\mu\nu}$, the curvature 
\be
\Omega_{\mu\nu} = [\nabla_{\mu},\nabla_{\nu}] = \partial_{\mu}\omega_{\nu} - \partial_{\nu}\omega_{\mu} 
+ [\omega_{\mu},\omega_{\nu}],
\ee
the Riemann  tensor $R_{\mu\nu\xi\lambda}$,  and also from covariant derivatives of all the objects mentioned above. For example $a_4^{\mathcal{M}}$  involves  the invariants 
$R^{\mu\nu\xi\lambda}R_{\mu\nu\xi\lambda}$, $\mathrm{tr} (\Omega_{\mu\nu}\Omega^{\mu\nu})$ and 
$\mathrm{tr} (Q_0 E_{;\mu}^{~~\mu})$,
 where ``$\mathrm{tr}$" stands for a trace over bundle indices. 
 Hereafter semicolon is a shorthand notation  for the covariant derivative with respect to Reimannian structure of $\mathcal{M}$, see the Appandix {\bf A}.
Note, that for odd values of $k$ all the bulk densities vanish identically. 

Boundary densities $a^{\partial\mathcal{M}}_k (\tilde{x},Q_0,L,\mathcal{B})$ are given by  linear combinations of local  boundary geometric invariants of the canonical dimension $k-1$ constructed from the quantities, mentioned above in the bulk context (viz. $Q_0$, $E$, $\Omega_{\mu\nu}$,..), and the purely boundary data: the extrinsic curvature
of the boundary $K_{\mu\nu}$, the endomorphism $S$,  the projectors  $\Pi_{\pm}$ 
and  covariant  derivatives of these entries. For example, see\footnote{A slightly corrected version of this formula is presented in Eq.~(A.7) of \cite{Gparity}.} 
\cite{Marachevsky:2003zb}, 
\begin{eqnarray}
&&a_3(Q_0,L)=\frac 1{384(4\pi)^{(n-1)/2}} \int_{\bM} \dd \tilde{x} \sqrt{\tilde{g}}\, \mathrm{tr}\, \left[
Q_0\bigl( -24E +24\chi E \chi +48\chi E +48 E\chi \right. \nonumber\\
&&\qquad +16\chi R - 8\chi R_{jn}^{\ \ jn} -12\chi_{:j}\chi^{:j} +12\chi_{:j}^{\ \ j}
+192 S^2+96KS +(3+10\chi)K^2 \nonumber\\
&&\qquad \left. +(6-4\chi )K_{ij}K^{ij} \bigr) +Q_{0;n}(96S + (18-12\chi)K) +24\chi Q_{0;nn}
\right], \label{a3}
\end{eqnarray}
where $\chi = \Pi_{+} - \Pi_{-}$, $K = K_{j}^{~j}$, $R$ is a scalar curvature on $\mathcal{M}$, and the colon is a shorthand notation  for the covariant derivative with respect to Reimannian structure of $\partial\mathcal{M}$, see the Appandix {\bf A}. The components of the four-dimensional tensors, such as $R_{jn}^{\ \ jn}$ or $K_{ij}$, are calculated in Gaussian normal coordinates near the boundary, which are chosen in the following way: $x^j := \tilde{x}^j$, $j=1,.., n-1$, and $x^n$ is the normal geodesic coordinate, see the Appendix {\bf B} for details.

In conclusion we consider the situation, when instead of $Q_0$ one  deals with the first order matrix valued differential operator $Q_1 = q^{\mu}\nabla_{\mu}$, where the quantities $q^{\mu}$ are matrix valued functions. In this case
the heat kernel expansion takes a slightly different form~\cite{Ref15Gparity}:
\be
K(Q_1,L,\mathcal{B},t) \simeq \sum_{k=-1}^{\infty} t^{\frac{k-n}{2}} a_k(Q_1,L,\mathcal{B}).
\ee 
It is shown in \cite{Gparity} that the  coefficients $a_k(Q_1,L,\mathcal{B})$ can be expressed through the heat kernel coefficients, which come out from the  ``standard" heat trace~\eqref{hc0}:
\be
a_k(Q_1,L,\mathcal{B}) =\frac{\partial}{\partial z}\bigg|_{z=0} a_{k+2}(\mathbb{1},L(z),\mathcal{B}) - \frac{1}{2} a_{k}(q^{\mu}_{~;\mu},L,\mathcal{B}), \la{prescript}
\ee
where one-parametric family of Laplace-type operators $L(z)$ is obtained from \eqref{25} replacing 
$\omega_{\nu}$ by $\omega_{\nu}(z) = \omega_{\nu} + \frac{1}{2}z q_{\mu}$, $z\in\mathbb{R}$.

In the subsequent sections we introduce the parity anomaly and explain, how it is related to the heat kernel techniques. After that we will show that this object can be useful in order to obtain nontrivial constrains on weights of some geometric invariants, which enter in $a_5$.

\section{Parity anomaly and heat kernel coefficients.}
Following \cite{Gparity} we 
consider a Dirac fermionic field $\psi$ which lives inside a compact\footnote{This is a technical requirement, which allows to use the zeta-function regularisation without having any problems in the infrared. Nowadays also the infrared frontier 
is getting interest \cite{Strominger:2017zoo,L1,L2}.
The parity-anomaly for a  non compact 4D manifold with boundary  was considered in~\cite{ChS} using different techniques.  } four-dimensional Riemannian manifold $\mathcal{M}$ with a boundary $\partial\mathcal{M}$. The boundary
 consists of several disjoint pieces, labeled by the subscript $\alpha$: $\partial\mathcal{M} = \cup_{\alpha} \partial\mathcal{M}_{\alpha}$. 
The classical dynamics is governed by the Dirac equation $\slashed{D}\psi = 0$, where the Dirac operator is given by 
\begin{equation}
\slashed{D}=\ii \gamma^a e_a^\mu\nabla_\mu, \qquad\mbox{with}\qquad
 \nabla_\mu = \partial_\mu+\tfrac 18 \sigma_{\mu ab}[\gamma^a,\gamma^b] \,.\label{Dop3}
\end{equation}
In this formula $\gamma^{a}$, $a = 1,..,4$ denote gamma matrices, which satisfy
\be
\{\gamma^{a}, \gamma^{b}\} = 2 \delta^{a b}\mathbb{1},
\ee
$e^a_{\mu}$ stand for the vielbeins, which obey
\be
e^a_{\mu}\,e^b_{\nu} \,\delta_{ab} = g_{\mu\nu},
\ee
and 
\be
\sigma_{\mu ab}=\Gamma_{\mu\nu}^\rho e_{\rho a}e^\nu_b - e^\nu_b\partial_\mu e_{\nu a}, \la{spin-connection}
\ee
is the spin-connection.

The local bag boundary conditions, which insure a vanishing current through the boundary 
\be
(\psi^{\dagger}\gamma^n \psi )\big|_{\partial\mathcal{M}} = 0, \qquad \gamma^{n}  \equiv \gamma^a\, e_{a}^{\mu}\, n_{\mu},
\ee
are chosen as follows:
\begin{equation}
\Pi_-\psi \vert_{\bM}=0,\qquad \Pi_-=\tfrac 12 (1-\ii \varepsilon_\alpha \gamma^5\gamma^n)\,, \qquad \varepsilon_\alpha \in \left\{-1,+1\right\},
\label{1stbc}
\ee
where $\gamma^5 = \gamma^1\gamma^2\gamma^3\gamma^4$.

Consider a quantum effective action
\begin{equation}
W_s(\slashed{D})=-\ln \det (\slashed{D})_s=\mu^s\Gamma(s)\zeta(s,\slashed{D})\,,\label{Ws}
\end{equation}
where we imposed the $\zeta$-function regularisation with
\begin{equation}
\zeta(s,\slashed{D})=\sum_{\lambda>0}\lambda^{-s} +e^{\ii \pi s}\sum_{\lambda<0}(-\lambda)^{-s}\,.
\label{zetaD}
\end{equation}
In this formula $s$ is the regulating parameter, and the physical limit corresponds to $s\longrightarrow 0$. Even though the classical Dirac equation is invariant with respect to the reflection
$\slashed{D}\longrightarrow -\slashed{D}$ of the Dirac operator, the quantum effective action is not: its reflection-odd part
\be
W^{\mathrm{odd}} := \lim_{s\longrightarrow 0}\frac{1}{2}\left( W_s(\slashed{D}) - W_s(-\slashed{D})\right)
\ee
is  different from zero. The nontriviality of $W^{\mathrm{odd}}$ is considered in \cite{Parity} as a generalisation of the parity anomaly for manifolds of an arbitrary dimension\footnote{In the odd-dimensional case this definition is in agreement with the notion of the parity anomaly, considered in 1980s \cite{Niemi:1983rq,Redlich:1983dv}. An interesting discussion on various definitions of the parity transformations can be found in~\cite{Novikov:2019kit} in the context of the pseudo-Hermitian PT-symmetric QFT. For the recent progress in this area 
we refer to~\cite{Novikov:2019ntb,Novikov:2018ybq,Novikov:2017dsl,Andrianov:2016ffj} and references therein.}. 

It is well known that $W^{\mathrm{odd}}$ can be expressed in terms of the spectral $\eta$-function \cite{AG}:
\be
W^{\mathrm{odd}}  = \frac{\ii\pi}{2}\,\eta(0,\slashed{D}), \la{PAnomEta}
\ee
where by definition
\begin{equation}
\eta(s,\slashed{D})=\sum_{\lambda>0}\lambda^{-s}-\sum_{\lambda<0}(-\lambda)^{-s}. \label{etas}
\end{equation}

From now on we use the Gaussian normal coordinates near $\partial\mathcal{M}$, see the Appendix {\bf B} for details. In the vicinity of the boundary the vielbeins can be chosen in the following way: $e_A$, $A=1,..,3$ is the orthonormal frame of the hypersurface $x^n = \mathrm{const}$, whilst
$e_n$ is the unit inward pointing normal to this hypersurface.

Let us perform the variation of vielbeins $e_A^{j}\longrightarrow e_A^{j}+\delta e_A^{j}$, which leaves the boundary condition \eqref{1stbc} invariant. The corresponding variation of the Dirac operator consists of two parts:
\be
\delta\slashed{D} = \mathcal{Q}_1 + \mathcal{Q}_0, \la{deltaD}
\ee
where
\be
\mathcal{Q}_1 = \ii (\delta e_A^{j} ) \gamma^A \nabla_{j}, \la{tdeltaD}
\ee
is the matrix valued first order differential operator, and
\be
\mathcal{Q}_0 = \frac{\ii}{8}e^{\mu}_a\gamma^a\left[\gamma_b,\gamma_c\right]\delta \sigma_{\mu }^{~bc},
\la{ttdeltaD}
\ee
is the matrix valued function.

The fundamental result is the following \cite{Gparity}: the outcoming variation of $W^{\mathrm{odd}}$ can be expressed in terms of
the heat kernel coefficient
\be 
\delta W^{\mathrm{odd}} = -\ii \sqrt{\pi}\, a_{3}(\delta\slashed{D}, \slashed{D}^2,\mathcal{B}), \la{parity-hc}
\ee 
where the boundary operator
$\mathcal{B}$ is given by Eq.~\eqref{gbc} with $\Pi_{\pm} =\tfrac 12 (1\pm\ii \varepsilon_\alpha \gamma^5\gamma^n)$ and $S=-\tfrac 12  K$.
At this point we recall the well-known fact \cite{manual}:  $L= \slashed{D}^2$ is the Laplace-type operator, which has the structure \eqref{25} with $\omega_{\nu} = \tfrac 18 \sigma_{\nu ab}[\gamma^a,\gamma^b]$ and $E = -R/4$. 

\section{Nontrivial constrains from the parity anomaly.}
\subsection{The idea.}
The 9 independent variations of the vielbeins $\delta e_{jA}$, $j=1,..,3$, $A=1,..,3$  are naturally divided into two groups.
The first
group can be parametrised via 6 independent real parameters $u_{AB} = u_{BA}$,
\be
\delta e_{jA} =  u_{AB} e^B_j. \la{svar}
\ee
Since $\delta g_{ij} = 2 u_{ij}$, these variations correspond to the variations of the metric tensor of the hypersurface $x^n = \mathrm{const}$. 
The remaining variations can be parametrised by  3 real parameters $v_{AB} = - v_{BA}$
\be
\delta e_{jA} = v_{AB}e^B_j, \la{avar}
\ee
and these variations correspond to  infinitesimal $SO(3)$ rotations of the local basis $e_A$, $A=1,..3,$ of the tangent space of the hypersurface $x^n = \mathrm{const}$. An $SO(3)$  rotation of the vielbeins 
\be
e_{A} \longrightarrow \hat{R}_{AB}\, e_{B} , \qquad \hat{R}\in SO(3),
\ee
generates a similarity transformation of the Dirac operator \eqref{Dop3}:
\be
\slashed{D} \longrightarrow \hat{S}\,\slashed{D}\,\hat{S}^{-1}, \qquad  \hat{S}\in Spin(4).
\ee
These transformations do not alter the spectrum
of the Dirac operator, therefore all the spectral functions, in particular the $\eta$-invariant \eqref{PAnomEta}, which defines the parity anomaly,
remain unchanged. 
In the original calculation of the gravitational parity anomaly~\cite{Gparity}, just the variations of the structure \eqref{avar} were considered, exactly for this reason.

In the present paper we redo the calculation of the gravitational parity anomaly, using smaller input:  
throughout the calculation we use just the general structure of the heat kernel coefficient $a_5$, where various geometric invariants enter with the \emph{unknown} weights. 
A lack of information will be compensated by the bigger group of allowed variations of the vielbeins:
apart from the variations \eqref{svar} we consider the infinitesimal rotations \eqref{avar} as well, requiring, however, that the variations of the $\eta$-invariant \eqref{PAnomEta} vanish under these infinitesimal rotations.

As we will see below, the last requirement appears to be sufficiently strong to fix \emph{all} the unknown weights of the nonzero
terms, which come out from $a_5$ during this calculation. 
  
\subsection{Relevant heat kernel coefficients.}
The heat kernel coefficient $a_{3}$, which enters in Eq.~\eqref{parity-hc}, naturally splits into two parts, c.f. Eq.~\eqref{deltaD}: 
\be
a_3 (\delta \slashed{D},\slashed{D}^2,\mathcal{B}) =
a_3 (\mathcal{Q}_1,\slashed{D}^2,\mathcal{B}) + a_3 (\mathcal{Q}_0,\slashed{D}^2,\mathcal{B}).
\la{a3gens}
\ee
In the Appendix {\bf C} we demonstrate that the second term of this sum is given by:
\bea
a_3 (\mathcal{Q}_0,\slashed{D}^2,\mathcal{B}) &=&-\frac{1}{32}\, \frac{1}{(4\pi)^{3/2}} { \int_{\partial\mathcal{M}} }
 \dd\tilde{x}\, \sqrt{g}\,\varepsilon_{\alpha}\,\left[ 
- \frac{2}{3} \,
  \tilde{R}_{:k} \, \epsilon^{nA~\,k}_{~~~q} \left(\delta e^q_A \right) -K_q^s K^r_{p:r} \left(\delta g_{sj}\right)  \epsilon^{njqp}
 \right. \nonumber \\
 &+& \left( 
K_{:j} K^{j}_k \epsilon^{nA~\,k}_{~~~q}
-K_{rp:k}K^{rp}\,\epsilon^{nA~\,k}_{~~~q}
- 2 K^A_p K^r_{k:r}\epsilon^{n~pk}_{~\,q} 
\right)\left(\delta e^q_A\right)  \nonumber\\
 &+& \left.  \left( 
   K_{:p}  - K^r_{p:r}
\right)\epsilon^{nA~\,p}_{~~~q} \left(\delta e^q_A\right)_{;n}
\right]. 
\la{a3c2final}
\eea
The first term of \eqref{a3gens} can be calculated with the help of the prescription \eqref{prescript} at $q^{j} =\ii (\delta e_A^{j} ) \gamma^A $,
and $q^n = 0$. Using the general formula \eqref{a3c2}, one can easily see that: 
\be
a_3 (q^{\mu}_{~;\mu},\slashed{D}^2,\mathcal{B})  =  0,
\ee
therefore 
\be
a_3 (\mathcal{Q}_1,\slashed{D}^2,\mathcal{B})  =\frac{\partial}{\partial z}\bigg|_{z=0} a_{5}(\mathbb{1},L(z),\mathcal{B}).   \la{prescript2}
\ee 
At this point the 5-th heat kernel coefficient enters in this game. Its general structure reads~\cite{a5}:
\begin{eqnarray}
&&a_5(f\cdot\mathbb{1},L,\mathcal{B}) =\frac{1}{5760(4\pi)^{(n-1)/2}} \int_{\partial\mathcal{M}} \dd \tilde{x}\,\sqrt{\tilde{g}} \,\mathrm{tr}\,\Big({w_{6} \,\chi \, \Omega_{jk}\Omega^{jk}} + {{w_{9}} \,\chi \, \Omega_{jn}\Omega^{jn}}  
\nonumber\\  
&&\qquad {w_{14} \,\chi \,\chi^{:j} \chi^{:p} \,\Omega_{jp} }  + \dots \Big),
\label{a5}
\end{eqnarray}
where $f$ is a smooth function, and ``$\dots$" stand for the terms, whose contribution to the right-hand side of \eqref{prescript2} equals to zero\footnote{Each of the unwritten terms ``$\dots$" vanishes separately.  We emphasise, that we consider \emph{all} the weights of all the geometric invariants, which enter in the general structure of the coefficient $a_5$, as unknown numbers. }. 

Repeating the discussion of~\cite{Gparity}, and implementing minor changes where needed, one arrives to the following answer:
\bea
a_3 (\mathcal{Q}_1,\slashed{D}^2,\mathcal{B})  &=& -\frac{1}{(4\pi)^{\frac{3}{2}}}\cdot\frac{1}{2880}  
 \int_{\partial\mathcal{M}}  \dd\tilde{x}\, \sqrt{g}\,\varepsilon_{\alpha}\,\Big[
w_6\left(\delta e^q_A \right)
  \tilde{R}_{BCq~\,:k}^{~~~~~k}    - w_9 \left(\delta e^q_A\right)_{;n} K_{qB:C}  \nonumber\\
&+&  w_9 \left(\delta e^q_A \right)K_{jq}K^{j}_{B:C}  
+2\left(w_6 - w_{14} \right)\left(\delta e^q_A \right)\big(K^j_{B}K_{qC} \big) _{:j}
\Big]\epsilon^{nABC} .  \la{a3c1}
\eea
We do not present  any calculational details here, since at this point the computations are almost identical to the ones
of~\cite{Gparity}.
\subsection{SO(3)-invariance.}
Since, as we said above, the P-odd part of the effective action \eqref{PAnomEta} remains unchanged upon the infinitesimal $SO(3)$ rotations \eqref{avar} of the
vielbeins, the relation \eqref{parity-hc} implies the following identity for the heat kernel coefficient \eqref{a3gens}:
\be
a_3 ({{\delta}} \slashed{D},\slashed{D}^2,\mathcal{B})\big|_{u=0} = 0. \la{mb}
\ee
On the other hand, a direct calculation, presented in the Appendix {\bf D} leads to:
\bea
a_3 ({{\delta}} \slashed{D},\slashed{D}^2,\mathcal{B})\big|_{u=0} &=& -\frac{1}{(4\pi)^{\frac{3}{2}}}\cdot\frac{1}{2880}  
 \int_{\partial\mathcal{M}}  \dd\tilde{x}\, \sqrt{g}\,\varepsilon_{\alpha}\,\bigg\{\Big[ \Big(\frac{w_6}{2} - 60\Big)\tilde{R}_{:k}
 + \Big(\frac{w_9}{2} - 90\Big)K^j_r K^r_{j:k} \nonumber\\
 &+& \Big(-\frac{w_9}{2} + w_6 - w_{14}\Big) K^j_r K^r_{k:j} + \Big(-90+w_6 - w_{14}\Big) K^r_{j:r} K^j_k \nonumber\\
 &+& \Big(90 - w_6 + w_{14}\Big)  K^r_{k:r} K +  \Big(90 - w_6 + w_{14}\Big)  K^r_{k} K_{:r} \Big] v_{AB} \nonumber\\
 &+& \Big(90 -\frac{w_9}{2}\Big)\left(K_{:k} - K_{k:r}^r\right)(v_{AB})_{;n} \bigg\}\epsilon^{nABk}. \la{a3ansW}
\eea
The first term of Eq.~\eqref{a3ansW} is the only term, which depends on the intrinsic geometry of the boundary but not on the 
extrinsic curvature, therefore the equality
\be
w_6 = 120 \la{w6}
\ee
is necessary for Eq.~\eqref{mb}. The last term of Eq.~\eqref{a3ansW} is the only term, which involves the normal derivatives of
the parameter $v$, therefore the condition
\be
w_9 = 180 \la{w9}
\ee
is necessary for Eq.~\eqref{mb} as well. Substituting the results \eqref{w6} and \eqref{w9} in \eqref{a3ansW}, we see, that the remaining terms vanish iff
\be
w_{14} = 30. \la{w14}
\ee
The results \eqref{w6}, \eqref{w9}, \eqref{w14} accompanied by the technique, which was used to obtain them, exhibit 
the main achievement of this paper.  

\subsection{What about the parity anomaly?}
It is remarkable, that the calculation presented above fixes \emph{all} the unknown constants in Eq.~\eqref{a3c1}, therefore
\be
a_3 ({{\delta}} \slashed{D},\slashed{D}^2,\mathcal{B}) = a_3 ({{\delta}} \slashed{D},\slashed{D}^2,\mathcal{B})\big|_{v=0}. 
\ee
The object, which stands in the right-hand side is exactly the one, which has been elaborated in \cite{Gparity}, so at this point no new computations are needed.

\section{Summary}
We have seen that the gravitational parity anomaly can be calculated  keeping
all the weights of various geometric invariants, which enter in $a_5$ as unknown numbers. 
The fact that the P-odd effective action, being actually the $\eta$-invariant, remains unchanged under local rotations of the vielbeins, 
allowed us to calculate all the relevant unknowns. This result suggests that  the parity anomaly might be a powerfool tool, which allows to generate nontrivial constrains 
on general structures of the odd heat kernel coefficients, which are allowed by symmetry and dimensional analysis. 

In conclusion, we acknowledge an interesting historical fact. In the original study of the fifth heat-kernel coefficient~\cite{a5} there was an 
error in the weight $w_{14}$, which was later corrected in~\cite{Moss:2012dp}. On the one hand the logic of~\cite{Moss:2012dp} relies on the fact that the weight $w_9 = 180$ was computed in~\cite{a5} correctly. On the other hand  the present consideration does not rely on any weight in the general structure \eqref{a5} at all, in particular the relation $w_9 = 180$ comes out directly from our calculations. 
From this point of view our study provides an efficient and independent crosscheck and confirmation of the correction, suggested in~\cite{Moss:2012dp}.

\section*{Appendix A: Summary of the notations.}
Throughout this article we use the following notations: 
\begin{itemize}
\item{the Greek letters $\mu, \nu, \rho, \sigma, \lambda, \eta$ label the 4-dimensional local coordinates $x$ on $\mathcal{M}$,
}
\item{ the small Latin letters $i,j,k,l,p,r,q$ numerate the 3-dimensional local coordinates $\tilde{x}$ on  $\partial\mathcal{M}$,}
\item{the small Latin letters $a,b,c,d,e,f$ label 4-dimensional vielbeins on $\mathcal{M}$,}
\item{ the capital Latin letters $A,B,C,D,E,F$ label 3-dimensional vielbeins on $\partial\mathcal{M}$. 
}
\end{itemize}
The vielbeins allow to ``translate" world  and flat indices into each other, e.g.
$
R^{\mu\nu ~ b}_{~~a~} = e^{\rho}_{a}e^{b}_{\sigma} R^{\mu\nu ~ \sigma}_{~~\rho~}.
$
We use tilde in order to indicate the  three-dimensional quantities, which are related to the intrinsic geometry of $\partial\mathcal{M}$,  e.g.  $ \tilde{R}^{i}_{~jkl}$ and $\tilde{\sigma}_{j ~B}^{~A~}$ stand for the Riemann tensor and the spin-connection on $\partial\mathcal{M}$ correspondingly.

The covariant derivatives with respect to the Riemannian structures of $\mathcal{M}$ and $\partial\mathcal{M}$ are denoted through the semicolon and through the colon respectively: 
 \bea
 (...)_{;\mu_1\,... \,\mu_p}^{~~~~~~~~\,\nu_1\,...\, \nu_r} &=& \nabla^{\nu_r}\,...\,\nabla^{\nu_1}\nabla_{\mu_p}\,...\,\nabla_{\mu_1}(...), \nonumber\\
 (...)_{: j_1\,... \,j_p}^{~~~~~~~~\,q_1\,...\, q_r} &=& \nabla^{q_r}\,...\,\nabla^{q_1}\nabla_{j_p}\,...\,\nabla_{j_1}(...).
 \la{shder}
 \eea
In order to avoid confusions we stress that apart from the corresponding Christoffel symbols both covariant derivatives involve the connection $\omega_{\mu}$ of the bundle $V$, which acts on endomorphisms $(...)$ of $V$ or $V|_{\partial\mathcal{M}}$ as a commutator $[\omega_{\mu}, (...)]$.

\section*{Appendix B: Gaussian normal coordinates.} 
In the vicinity of a given component $\partial\mathcal{M}_{\alpha}$ of the boundary,
 the Gaussian normal coordinates can be constructed as follows: the first $n-1$ coordinates coincide with the coordinates on $\partial\mathcal{M}_{\alpha}$, i.e. $x^j:= \tilde{x}^{j}$, whilst  $x^n$ is the geodesic normal coordinate.
In this coordinate system the infinitesimal interval $ds^2 = g_{\mu\nu}dx^{\mu}dx^{\nu}$ takes a simple form
\be
ds^2 = dx^n dx^n  + g_{jk} d\tilde{x}^j d\tilde{x}^k,
\ee
and the determinants  of the 4d and the 3d metric tensors coincide: $\tilde{g} = g$.
By definition we suppose that the boundary is located at $x^n = 0$.
The three-dimensional metric, vielbeins, Christoffel symbols and spin-connection coincide with the proper components of the corresponding
4-dimensional objects:
\bea
\tilde{g}_{jk} &=& g_{jk}, \quad \tilde{g}^{jk} = g^{jk}, \nonumber \\
\tilde{e}_{A}^{j} &=& e_A^j, \quad \tilde{e}^{A}_{j} = e^A_j, \nonumber \\
\tilde{\Gamma}^{i}_{jk} &=& {\Gamma}^{i}_{jk}, \nonumber \\
\tilde{\sigma}_{j ~B}^{~A~} &=& {\sigma}_{j ~ B}^{~A~}.
\eea  
If one of the 4-dimansional indeces equals to $n$, the following relations hold:
\bea
g_{nn} &=& g^{nn} = +1, \quad g_{jn} = g_{jn} = g^{jn} = g^{nj} = 0, \nonumber\\
e_{n}^{n} &=& +1, \quad e_j^n = e_{n}^j = e_n^{A} = e_A^n = 0, \nonumber \\
\Gamma^n_{jk} &=& K_{jk}, \quad \Gamma^j_{nk} = \Gamma^j_{kn} = - K^j_k, \quad 
\Gamma^{n}_{nn} = \Gamma^j_{nn} = \Gamma^n_{jn} = \Gamma^n_{nj} = 0. \la{GNCfacts}
\eea
In the Gaussian normal coordinates the nonzero components of the extrinsic curvature read\footnote{All other components of the extrinsic curvature viz $K_{nn}$ and $K_{jn} = K_{nj}$ vanish in this coordinate system.}:
\be
K_{jk} \equiv -\frac{1}{2}\partial_n g_{jk}, \la{KGNC}
\ee  
{and the Ricci equation\footnote{At this point we use the terminology of~\cite{Gourgoulhon:2007ue}. Note, that in the Gaussian normal coordinates Eq.~(3.43) of this reference becomes much simpler, in particular, a contribution of the lapse function vanishes.} reduces to:
\be
R_{nqnp}= \partial_{n} K_{qp} + K^r_q K_{pr}.   \la{Mainardi}
\ee
Strictly speaking the extrinsic curvature is defined on the boundary $x^n = 0$ only, therefore the meaning of its normal derivative requires clarifications. At $x^n\neq 0$ the quantity $K_{qp}(\tilde{x}, x^n)$ is naturally identified with the extrinsic curvature of the hypersurface $x^n = \mathrm{const}$, i.e. with $\Gamma^n_{qp}$, therefore in the sufficiently small vicinity of the boundary, where the Gaussian normal coordinates are well-defined, the normal derivative $ \partial_{n} K_{qp}$ makes perfect sense.
 
The quantities $R^{i}_{~jkl} $ and $ \tilde{R}^{i}_{~jkl}$  are connected via the Gauss equation:
\be
R^{i}_{~jkl} = \tilde{R}^{i}_{~jkl} + K_{jk}K^i_l - K_{jl}K^i_k, \la{Gauss} 
\ee
while the Codazzi equation  expresses $R^{n}_{~ jkl}$ through the extrinsic curvature:
\be
R^{n}_{~jkl} = K_{jl:k} - K_{jk:l}. \la{Codazzi} 
\ee
} 
\section*{Appendix C: derivation of Eq.~\eqref{a3c2final}.}
Throughout this Appendix we work with the Gaussian normal coordinates near the boundary.   
Using the general formula
\eqref{a3} at $Q_0 = \mathcal{Q}_0$ (see the definition in Eq.~\eqref{ttdeltaD}), one can easily chek that:
\bea
a_3 (\mathcal{Q}_0,\slashed{D}^2,\mathcal{B}) &=& -\frac{1}{384}\frac{1}{(4\pi)^{3/2}}
{\int_{\partial\mathcal{M}}} \dd \tilde{x}\, \sqrt{g}\,\varepsilon_{\alpha}\,\left[\left( - 8R - 22 K^r_q K^q_r + 10 K^2  {-} 8 
R^{q}_{~nqn} \right)\mathcal{F}_1 \right. \nonumber\\ 
&+& \left. 12 \,{\mathcal{F}}_2 + 
12\, \hat{\mathcal{F}}_3 \right], \la{a3c2}
\eea
where by definition:
\bea
\mathcal{F}_1 &:=& -\ii \cdot\mathrm{tr}\,\left(\mathcal{Q}_0\,\gamma^5\gamma^{\mu}\,n_{\mu}\right)
= e^j_A (\delta\sigma_{j BC} ) \epsilon^{nABC}, \la{F1def} \\
{\mathcal{F}}_2 &:=& K^r_{D:r}e^{\mu}_a  (\delta \sigma_{{\mu}bc}) \epsilon^{abc D},  \la{F2def} \\
\hat{\mathcal{F}}_3 &:=&
 - 3 K ({\mathcal{F}}_1)_{;n} + 2 ({\mathcal{F}}_1)_{;nn}.
\la{S3def}
\eea
\emph{Comment. Strictly speaking, the unit normal $n_{\mu}$ is defined on the boundary only, and so is $\mathcal{F}_1$. Let us clarify 
what the normal derivatives of $\mathcal{F}_1$  
mean. For each component $\partial\mathcal{M}_{\alpha}$ of the boundary there exists a sufficiently small $\delta > 0$, such that for any $x^n\in [0;\delta]$ one can  build a  hypersurface  $x^n = \mathrm{const}$.
Identifying $n_{\mu}$ with the unit normal to this hypersurface, we can naturally extend the definition \eqref{F1def} to a vicinity $\partial\mathcal{M}_{\alpha}\times [0;\delta]$ of the boundary. From now on the normal derivatives of  $\mathcal{F}_1$ refer exactly to this extended definition.}

The variations of the spin-connection \eqref{spin-connection} read:
\bea
\delta\sigma_{j AB} &=&  
\frac{1}{2} e^r_A e^q_B \left(-\left(\delta g_{jq}\right)_{:r} + \left(\delta g_{jr}\right)_{:q} + \left(\delta g_{rq}\right)_{:j}\right)
-e^q_B \left(\delta e_{Aq}\right)_{:j} \la{varLCjAB}
, \\
\delta\sigma_{nAB} &=& \frac{1}{2}\, e^{j}_A e^q_B \left(
K^s_q \left(\delta g_{sj}\right) - K^s_j \left(\delta g_{sq}\right)  
+\left(\delta g_{qj}\right)_{;n} \right)
-e^q_B \left(\delta e_{Aq}\right)_{;n}, \la{varLCnAB} \\
\delta\sigma_{qAn} &=& \frac{1}{2} e^j_A\left(
K_q^s \left(\delta g_{sj}\right)  - K_j^s \left(\delta g_{qs}\right) + \left(\delta g_{jq}\right)_{;n}
\right) - K^j_q \left(\delta e_{Aj}\right). \la{varLCqAn}
\eea
Using the relation \eqref{varLCjAB} we rewrite  Eq.~\eqref{F1def} 
as follows:
\bea
{\mathcal{F}}_1 
= (e^j_A e_{Bq} (\delta e^q_C) \epsilon^{nABC})_{:j}. \la{comb1}
\eea
Lengthy but otherwise straightforward computations, based on the equalities \eqref{varLCnAB} and \eqref{varLCqAn}, demonstrate 
that:
\be
{\mathcal{F}}_2 =- K_q^s K^r_{C:r} \left(\delta g_{sj}\right)  \epsilon^{njqC} - K^r_{C:r}\left(\delta e_{A}^q\right)_{;n} \epsilon^{nA~\,C}_{~~~\,q}
+ 2\left(\delta e^q_{A}\right) K^A_B  K^r_{C:r} \, \epsilon^{nB~\,C}_{~~~\,q}. \la{comb2}
\ee

Now we focus on a contribution of the third (and the last) combination $\hat{\mathcal{F}}_3$ to Eq.~\eqref{a3c2}. 
The relation 
\be
\int_{{\partial\mathcal{M}_{\alpha}}} \dd \tilde{x} \sqrt{g}\, {(\mathcal{F}_1)}_{;nn} = \int_{{\partial\mathcal{M}_{\alpha}}} 
\dd \tilde{x} \sqrt{g} \left(  
2K{(\mathcal{F}_1)}_{;n} + \,(\partial_n K)\,{\mathcal{F}_1} - K^2 {\mathcal{F}}\right). \la{usefulEq1}
\ee
allows to get rid of a contribution of the second normal derivative in \eqref{S3def}. \\
 
\noindent\emph{Proof of Eq.~\eqref{usefulEq1}. Being a total tangential derivative, the function ${\mathcal{F}}_1$ obviously satisfies the following property: 
\be
\int_{{\partial\mathcal{M}_{\alpha}}} \dd \tilde{x} \sqrt{g} \,{\mathcal{F}_1} = 0, { \quad\mbox{at}\quad \forall x^n \in [0,+\delta]}. \la{defprop}
\ee
At this point we refer to the extended definition of $\mathcal{F}_1$, discussed  above. 
Calculating the second normal derivative of Eq.~\eqref{defprop} at $x_n = 0$, and using the relation
\be
\partial_{n} \sqrt{g}  = -\sqrt{g}\, K  \la{detsqrtg}, 
\ee
we immediately get Eq.~\eqref{usefulEq1}.
}
\\

\noindent A straightforward computation leads us to:
\bea
\left({\mathcal{F}}_1\right)_{;n}  
 = \left(e^j_A e_{Bq} (\delta e^q_C)_{;n} \epsilon^{nABC}\right)_{:j}-K_{:A}  \left(\delta e^q_C \right) 
 \epsilon^{nA~\,C}_{~~~\,q} + \left(K^j_A \left(\delta e^q_C \right) \epsilon^{nA~\,C}_{~~~\,q}\right)_{:j}.
 \la{usefulEq2}
\eea
The results \eqref{usefulEq1} and \eqref{usefulEq2} imply that the combination 
$\hat{\mathcal{F}}_3$  can be replaced by the expression
\be
{\mathcal{F}}_3 = \left(K_{:C}\left(\delta e^q_A \right)_{;n} + 2K_{;n:C}\left(\delta e^q_A \right)\right)\epsilon^{nA~\,C}_{~~~\,q}
       +\left(  
       K_{:j}K^j_C - \frac{3}{2}\left(K^2\right)_{:C}
       \right)\left(\delta e^q_A \right)\epsilon^{nA~\,C}_{~~~\,q} \la{comb3}
\ee
under the integral sign in Eq.~\eqref{a3c2}.
Using this fact together with the  equations \eqref{comb1}, \eqref{comb2} 
and the relations\footnote{These relations are the consequences of  the Ricci \eqref{Mainardi}  and the Gauss \eqref{Gauss} equations. We remind, the meaning of the normal derivative of the extrinsic curvature in this context is clarified in the Appendix {\bf B}.}
\bea
R &=& \tilde{R} + 2\,\partial_n K - K_{rq}K^{rq} - K^2,
 \nonumber \\
R^{q}_{~nqn} &=& \,\partial_n K - K_{rq}K^{rq}, \la{Rsimp2}
\eea
one immediately obtains Eq.~\eqref{a3c2final}. 

\section*{Appendix D: derivation of Eq.~\eqref{a3ansW}.}
Substituting Eq.~\eqref{avar} in Eq.~\eqref{a3c2final} and in Eq.~\eqref{a3c1} we get
\bea
a_3 (\mathcal{Q}_0,\slashed{D}^2,\mathcal{B})\big|_{u=0} &=&-\frac{1}{32}\, \frac{1}{(4\pi)^{3/2}} { \int_{\partial\mathcal{M}} }
 \dd \tilde{x}\, \sqrt{g}\,\varepsilon_{\alpha}\,\Big[ 
 \,
  \Big(- \frac{2}{3}\tilde{R}_{:k} 
 + K_{:j} K^{j}_k 
-K_{rp:k}K^{rp}
  \Big)\, \epsilon^{nABk} v_{AB} 
  \nonumber \\
 &+& 
 \left( 
   K_{:p}  - K^r_{p:r}
\right)\epsilon^{n A B p} \left(v_{AB}\right)_{;n}
- 2 K^A_p K^r_{k:r}\epsilon^{nBpk}
v_{AB} \Big]
\la{a3c2finalA},
\eea
and
\bea
a_3 (\mathcal{Q}_1,\slashed{D}^2,\mathcal{B})\big|_{u=0}  &=& -\frac{1}{(4\pi)^{\frac{3}{2}}}\cdot\frac{1}{2880}  
 \int_{\partial\mathcal{M}}  \dd \tilde{x}\, \sqrt{g}\,\varepsilon_{\alpha}\,\Big[
w_6\, v_{AB}\,
  \tilde{R}_{DC~~~:k}^{~~~~Bk}    - w_9 \left(v_{AB}\right)_{;n} K^{B}_{D:C}  \nonumber\\
&+&  w_9 \, v_{AB}\,K_{j}^B K^{j}_{D:C}  
+2\left(w_6 - w_{14} \right)\, v_{AB}\,\big(K^j_{D}K_{C}^B \big) _{:j}
\Big]\epsilon^{nADC},   \la{a3c1A}
\eea
respectively.
Using the identity
\be
v_{AB}K^A_p K^r_{k:r}\epsilon^{nBpk} =\frac{1}{2} v_{AB}\epsilon^{nABk}(-K K^r_{k:r} + K^j_k K^r_{j:r} ) \la{i1}
\ee
and the equations
\bea
\tilde{R}_{DC~~~:k}^{~~~~Bk}\,\epsilon^{nADC} v_{AB} &=& \frac{1}{2}\,\tilde{R}_{:k}\, \epsilon^{nABk}v_{AB} \nonumber\\
(v_{AB})_{;n} K^{B}_{D:C}\epsilon^{nADC} &=& \frac{1}{2} (v_{AB})_{;n}\left(K_{:p} - K_{p:r}^r\right)\epsilon^{nABp} \nonumber\\
v_{AB}\,K_{j}^B K^{j}_{D:C}\epsilon^{nADC} &=& \frac{1}{2} v_{AB} \epsilon^{nABk}\left(K^j_r K^r_{j:k} - K^j_r K^r_{k:j}\right)
\nonumber\\
v_{AB}\,\big(K^j_{D}K_{C}^B \big) _{:j}
\epsilon^{nADC} &=&  \frac{1}{2} v_{AB} \epsilon^{nABk} ( K^r_j K^j_k - K^r_k K)_{:r} \la{i2}
\eea
we immediately arrive to Eq.~\eqref{a3ansW}.

In conclusion, for the sake of completeness, we clarify the origin of the relations \eqref{i1} and \eqref{i2}.
First we notice that an arbitrary 3d anti-symmetric tensor $v_{AB}$ satisfies
\be
v_{AB}\epsilon^{nADC} = \frac{1}{2} v_{EF} \left(\epsilon^{nCEF}\delta^{D}_B - \epsilon^{nDEF}\delta^{C}_B\right). \la{i3}
\ee
Indeed,  
\be
v_{AB} = \frac{1}{2}v_{EF}\left(\delta_{A}^{E}\delta_{B}^{F} - \delta_{B}^{E}\delta_{A}^{F}\right) 
= \frac{1}{2}\,v_{EF}\, \epsilon^{nGEF}\,\epsilon_{nGAB}, \la{i4}
\ee
where we used the well-known rule for a contraction of 3d Levi-Civita tensors.
Substituting Eq.~\eqref{i4} in the left-hand side of Eq.~\eqref{i3} and applying this rule for the combination $\epsilon_{nGAB}\epsilon^{nADC}$  we obtain the right-hand side of Eq.~\eqref{i3}.

Applying the identity \eqref{i3} to the left-hand sides of Eq.~\eqref{i1} and Eq.~\eqref{i2}, 
one can easily see a validity of these equations. In order to derive the first equation of \eqref{i2} one 
has to use the contracted Bianchi identity
\be
\tilde{R}_{c:k}^{k} = \frac{1}{2}\tilde{R}_{:c}.
\ee
\section*{Acknowledgments}
MK is grateful to Dmitri Vassilevich for collaboration on the basic papers \cite{Parity} and \cite{Gparity} and related topics.  MK is supported by INFN Iniziativa Specifica GeoSymQFT.  

\end{document}